\documentclass[superscriptaddress,floatfix,aps,nofootinbib]{revtex4}
\usepackage{epsfig}

\def\lp {\left( }
\def\rp {\right) }
\def\lb {\left[ }
\def\rb {\right] }
\def\lc {\left\{ }
\def\rc {\right\} }

\def\nn {\nonumber}

\def\beq{\begin{equation}}
\def\eeq{\end{equation}}
\def\bea{\begin{eqnarray}}
\def\eea{\end{eqnarray}}
\def\ni{\noindent}

\def\p{\pi}

\begin{document}

\title{$\Lambda$  and $\overline{\Lambda}$ polarization in Au-Au collisions at RHIC}

\author{C. C. Barros Jr.}
\affiliation{Depto de F\'{\i}sica - CFM - Universidade Federal de Santa
Catarina 
 Florian\'opolis - SC - CP. 476 - CEP 88.040 - 900 - Brazil}

\author{Y. Hama}
\affiliation{Instituto de F\'{\i}sica, Universidade de S\~{a}o Paulo,
C.P. 66318, 05315-970, S\~ao Paulo, SP, Brazil}




\begin{abstract}
Experiments at RHIC have shown that 
in 200 GeV Au-Au collisions, the
$\Lambda$ and $\overline{\Lambda}$ hyperons are
 produced  with very small polarizations
\cite{RH1}, almost consistent with zero.
  These results can be understood in terms of  
 a model that we proposed \cite{cy}. In this work, we show how this 
model may be applied in such collisions, 
and also will discuss the relation of our results with
 other models,
in order to explain the experimental data.
\end{abstract}

\maketitle

\vspace{5mm}

Since the discovery of significant polarization for the $\Lambda$ 
particles produced
in 100 GeV p-Be collisions by Bunce \cite{bu}, hyperon polarization has shown 
to be a very challenging subject, as, at the time it was a totally
surprising result. 
This fact, unexpected both experimentally and theoretically 
has been confirmed by further experiments, and this puzzle has been 
complicated
when the polarizations of the other hyperons and antihyperons have been 
measured \cite{hel}-\cite{mor}.

Hyperon polarization may be quite well described by parton-based models
\cite{mod1}-\cite{modn}, 
but antihyperon polarization not.
In  \cite{cy}, we proposed a model \cite{cy} that was able to describe 
successfully the 
antihyperon polarization in terms of final-state interactions
that occur in the hadronic phase of such collisions,
in a mechanism 
based in relativistic hydrodynamics.

Recently, at RHIC, in 200 GeV Au-Au collisions, the $\Lambda$ and 
$\overline{\Lambda}$ polarizations have been measured \cite{RH1}, as functions
of the transverse momentum, in the range $0<p_t <5$ GeV, and as functions of 
the 
pseudorapidity, in the range $-1.5<\eta<1.5$. In this region, the final 
polarization  for both particles
may be considered consistent 
with
zero. As it was suggested in \cite{pan}, zero polarization in high energy nucleus-nucleus interactions, if observed, could show a signal of quark-gluon plasma formation. Some models show good results in explaining $\Lambda$ and 
$\overline{\Lambda}$ polarizations.
 In  \cite{pj}-\cite{lw2}, this effect is proposed as the partons
are produced with large angular momentum, and 
quark polarization  results
 from parton scattering. In \cite{bec1}, \cite{bec2} polarization of spin 1/2 particles for an 
equilibrated system is computed.

As we can see, this is a very important problem,
 and the objective of this letter is
to study this question, showing some results that we obtained, and
discussing their relations with other theoretical results.
We will apply the model that we used to calculate 
antihyperon polarization in p-A collisions, in the study of the 
Au-Au collisions performed at RHIC. In \cite{cy}, we have shown that significant polarization may occur considering this model. 
We want to investigate the effect of the final-state 
interactions in nucleus-nucleus collisions and if it is possible that these interactions may affect the final polarization of the produced particles.
This model is based on the
hydrodynamical aspects of such collisions, so,
the first step is to obtain the 
velocity distribution of the fluid formed during the collision. Then, we will 
use it in order to obtain the average polarization, taking into account
the $\p\Lambda$ and  $\p\overline{\Lambda}$
final interactions.

In  the hydrodynamical picture,
in the collision of two high-energy particles, the large amount of energy 
localized in a very small volume produces a fluid, that expands and then 
produces the final particles, what may be understood by the freeze-out 
mechanism.
 We will suppose a parametrization of the velocity distribution of 
such fluid given by the expression
\beq
u^0 {d\rho\over d^3u}=A\lb e^{-\beta (\alpha-\alpha_0)^2}+
  e^{-\beta (\alpha+\alpha_0)^2} \rb e^{-\beta_t \xi^2}  \  \  ,
\label{flu}
\eeq

\ni
that is written
 in terms of its longitudinal ($\alpha$) and transversal ($\xi$)
rapidities.
That means that the formed fluid expands in the 
 the incident nuclei direction ($\alpha$), and also in the transverse
direction ($\xi$). This kind of velocity distribution has shown to describe
correctly the production of particles in many other systems \cite{yxt}, 
\cite{yf}.
We may visualize this fluid geometrically, in a first approximation, as an hot 
expanding cylinder.
The constants
 $\beta$, $\beta_t$ and $\alpha_0$ are parameters 
that describe the shape of this distribution, and are
determined by calculating the distributions of the produced particles,
and, comparing them with
 the RHIC experimental data for the transverse momentum
$p_t$ \cite{Rpt} and 
pseudorapidity  ($\eta$) distributions \cite{Ret}.

 This objective may be achieved, making
 a convolution of the fluid velocity distribution (\ref{flu}),
 with the particles 
distribution, inside these fluid elements, that may be considered a Bose
distribution as most of the produced particles are pions.
We will consider
\beq
{dN\over d\vec{p}_0} = {N_0\over{\exp{(E_0/ T)}}-1}
\eeq

\ni
with the temperature $T\sim m_\pi$, and
 $\vec p_0$ and $E_0$ are the momentum and energy of the pions inside one 
fluid element.
 So, the observed distributions of particles
are given by
\bea
E{dN\over d\vec{p}}=C\int \lb e^{-\beta (\alpha-\alpha_0)^2}+
  e^{-\beta (\alpha+\alpha_0)^2} \rb e^{-\beta_t \xi^2}  \\  \nn
\times    {E_0(\alpha,\xi,\phi)\over{\exp{(E_0(\alpha,\xi,\phi) /T)}}-1}
\rm{sinh}\xi \ \rm{cosh}\xi \ d\alpha \ d\xi
\  d\phi   \   \    ,
\label{dn}
\eea

\ni
where $\phi$ is the azimuthal angle.
The results of the particles distributions resulting from eq. (\ref{dn})
 are shown in Figures 1 and 2. We obtained a 
very good description of $d\sigma/d\eta$ for all centralities (Fig. 1), and 
for the $p_t$ distribution (Fig. 2), the results are very good for
$p_t< 6$ GeV. For $p_t> 6$ GeV, a small discrepancy may be noticed, and
it increases with $p_t$.
This fact is not a problem for the present work, as in the 
experimental data for polarization, the values of $p_t$ investigated are below
 this value. This problem shows that other processes become important at large $p_t$, such as the 
hard scattering ones. A way to improve the results is to insert
 an extra term, depending on the transverse
rapidity of the fluid $\xi$, in eq. (\ref{flu}), what represents alterations in the equation of state.
 For simplicity, it will not
be done in this paper.

\begin{figure}[hbtp]
\centerline{
\epsfxsize=85.mm
\epsffile{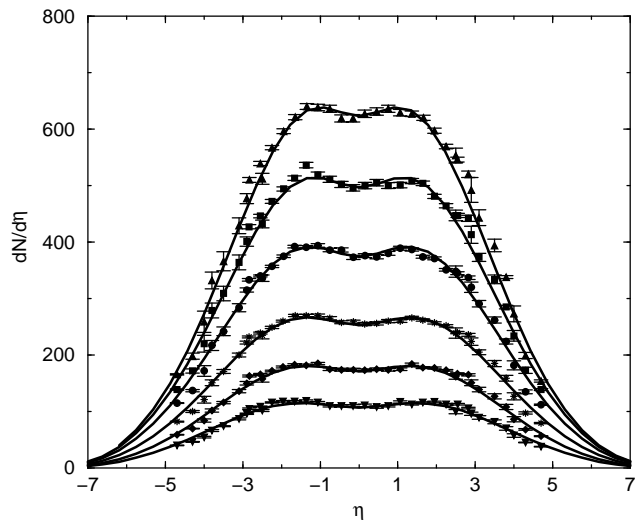}}
\caption{Distributions $dN/d\eta$, for many centralities. From the top,
 0-5\%, 5-10\%, 10-20\%, 20-30\%, 30-40\%, 40-50\%. 
We compare our results  (solid lines),
 with the experimental data from \cite{Ret} (points).}
\end{figure}

\begin{figure}[hbtp]
\centerline{
\epsfxsize=83.mm
\epsffile{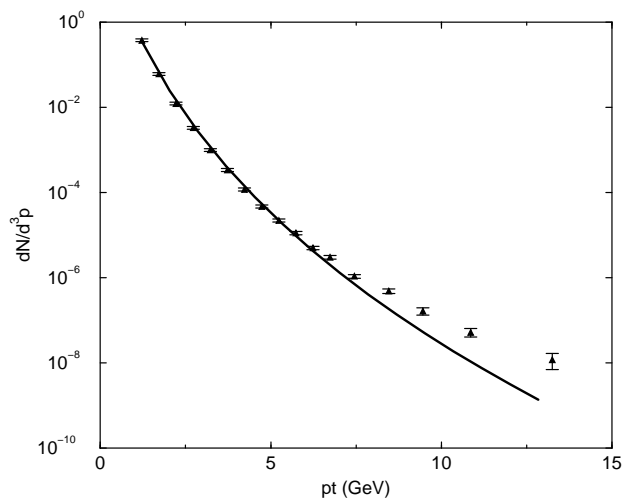}}
\caption{ Comparison of the calculated
distribution $dN/d^3p$ as function of $p_t$
 with the experimental data from \cite{Rpt}.}
\end{figure}

The parameters obtained are $\beta$=0.14, $\beta_t$=3.2 and $\alpha_0$ in
the range 1.5-1.75, varying with the centrality, as it can be seen in Table I.
We can observe that $\beta$ and $\beta_t$ does not seem to have any dependence
on the centrality.

\begin{table}[hbt]
\begin{center}
\caption{Values of the parameters $\beta$, $\beta_t$ and $\alpha_0$ 
of the curves shown Figs. 1 and 2} 
\begin{tabular} {|c|c|c|c|}
\hline

Centrality		& $\beta$ 	& $\beta_t$ 	
& $\alpha_0$		\\ \hline
0-5\%	& 0.14  & 3.2  & 1.50		\\  \hline
5-10\%  &  0.14 & 3.2  & 1.56 \\ \hline
10-20\% & 0.14  & 3.2  & 1.62  	\\ \hline
20-30\% & 0.14  & 3.2  & 1.67  	\\ \hline
30-40\% & 0.14  & 3.2  & 1.70  	\\ \hline
40-50\% & 0.14  & 3.2  & 1.75  	\\ \hline
\end{tabular}
\end{center}
\end{table}

Observing these results, one may see that the fluid parametrization,
with the velocities distributions given by (\ref{flu}), is very 
reasonable and describes quite well the experimental data in the region of our
interest. So, considering this description, we may calculate the polarization
of a hyperon (or antihyperon) produced in the interior of such system,
taking into account the effect of the final-state interactions, of these particles
with the surrounding pions (that is the dominant effect), as we made in 
\cite{cy}. 

Now, let us turn our attention to the final-state interactions. 
The most important case to be considered is the $\pi\Lambda$ 
($\pi\overline{\Lambda}$), as it is the most probable interaction.
The relative energy of this interaction is not so high, despite
the fact that these particles are observed with high energies in the
laboratory system of reference. This interaction may be described
by effective chiral lagrangians, as we made in \cite{BH}-\cite{ccb},
where the resonance $\Sigma^*$(1385) in the intermediate state is a key 
element. These lagrangians are

\begin{eqnarray}
{\cal{L}}_{\Lambda\pi\Sigma} &=& {g_{\Lambda\pi\Sigma}\over 2m_\Lambda}\lc
 {\overline\Sigma}
\gamma_\mu \gamma _{5} \vec \tau\Lambda \rc .\partial^\mu \vec \phi  
\ + h.c.
\\
{\cal{L}}_{\Lambda\pi\Sigma^*} &=& g_{\Lambda\pi\Sigma^*} \lc 
{{\overline{\Sigma}^*}^\mu}
\lb g_{\mu\nu} - \lp Z+{1\over 2}\rp \gamma_\mu \gamma _\nu  
\rb \vec\tau\Lambda \rc .\partial ^\nu \vec \phi \nn \\
&&
 + h.c.  \  \ , 
  \label{3.1} 
\end{eqnarray}

\ni
 where $\vec\phi$  is the pion field and $Z$ is a 
parameter representing the possibility of the off-shell-resonance having 
spin 1/2. 

The considerd diagrams for the scattering amplitude are shown in Figure 3. 
The scattering amplitude determines the cross sections $d\sigma/d\Omega$ and $d\sigma/dt$, and 
the polarization. More details on the calculations may be found in \cite{cy}, \cite{cm} and \cite{BH}. 
The  $\Lambda$ polarization, as a function of $x=\cos\theta$, where $\theta$ is the scattering angle, is shown in Figure 4. 

\begin{figure}[hbtp] 
\centerline{
\epsfxsize=8.cm
\epsffile{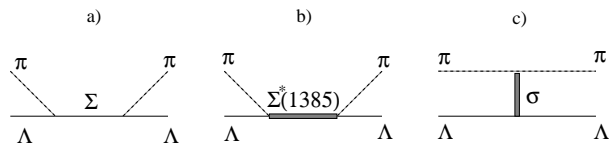}}
\label{f11}
\caption{ Diagrams for $\pi\Lambda$ Interaction}
\end{figure}

One must observe that this model, that  we proposed in 2001 \cite{BH}, has made a very
good prediction for the $\pi\Lambda$ phase shift at the $\Xi$ mass,
$\delta_P-\delta_S$=$4.3^o$, result that
 has been confirmed experimentally 
at the Fermilab in the HyperCP experiment in 2003 \cite{hyper1},\cite{hyper2},
where they obtained $\delta_S-\delta_p$=$(4.6\pm1.4\pm 1.2)^o$.
This result validates our model for the $\pi\Lambda$ interaction.

\begin{figure}[hbtp] 
\centerline{
\epsfxsize=8.cm
\epsffile{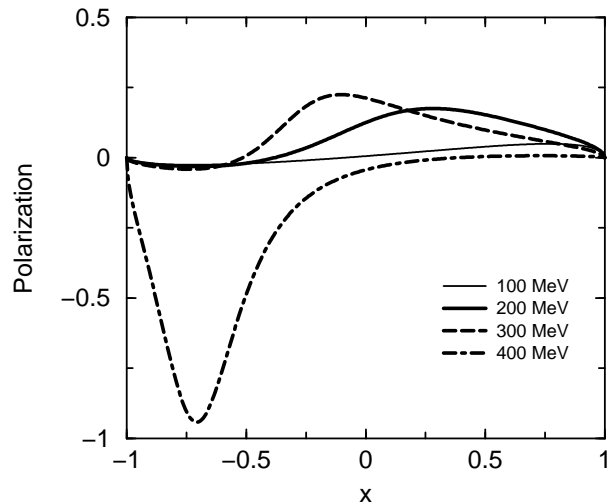}}
\caption{Polarization in the $\pi\Lambda$ interaction, $x=$cos$\theta$.}
\label{pilmpol}
\end{figure}

With the knowledge of the velocities distribution and of the final 
interactions, we are able to calculate
 the average polarization of the produced particles in the same way we made in
\cite{cy}.

 The average 
polarization may be calculated by the expression 
\beq
 \langle\vec P\rangle 
 ={\int \left(\vec P^\prime \ d\sigma / dt\right)
  {\cal{G}}  \  d\alpha\ d\xi d\phi \ d\vec \Lambda _0'd\vec\pi _0'
  \over\int \left({d\sigma / dt}\right)
  {\cal{G}} \  d\alpha\ d\xi d\phi \ d\vec \Lambda _0'd\vec\pi _0'   } \ \ ,
\label{pm}
\eeq

\ni 
where $\vec\Lambda_0'$ is the $\Lambda$ momentum and $\vec\pi_0'$ is the pion one. 
The factor ${\cal{G}}$ that appears in eq. (\ref{pm}) contains 
the statistical weights of the production of the particles and 
the ones relative to the expansion of the fluid, and can be 
written as
\bea
{\cal{G}}=&&{(d\rho /d^3u) \over 
\left(\exp\left(E_{\pi_0}'/ T\right)-1\right)
\left(\exp\left(E_0'/ T \right)+1\right)} 
\Lambda _0'^2\pi _0'^2 \nn \\
&&\times
\delta \left(E_0'+E_{\pi _0}'-E'-\sqrt {m_\pi ^2 + (\vec \pi_0'+
\vec \Lambda _0'-\vec \Lambda ')^2}\right)  
\  \   , \nn \\
\eea

\ni
where $d\rho /d^3u$, is given by (\ref{flu}).

With this procedure 
we obtained 
the results shown in Figs. 3 and 4. As we can see, the resulting polarization 
is very small (smaller than 1\%) for all values of the centrality,
 and are in good accord with the experimental 
data for both $\Lambda$ and $\overline{\Lambda}$.

\begin{figure}[hbtp]
\centerline{
\epsfxsize=85.mm
\epsffile{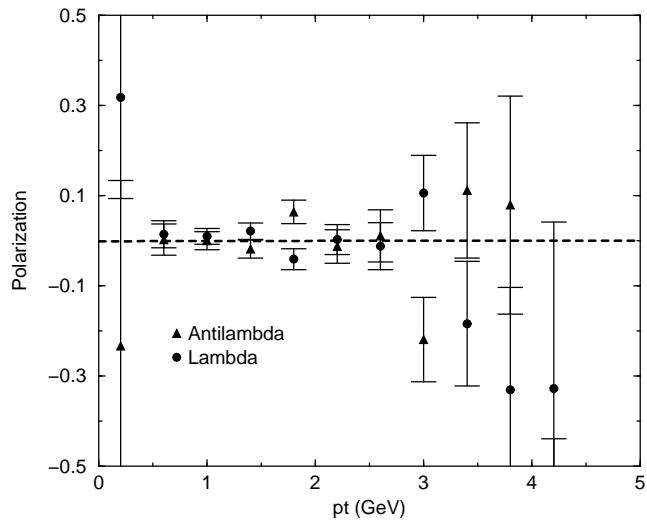}}
\caption{ Calculated polarization (dashed line)
 as function of $p_t$ compared 
with the experimental data from \cite{RH1}.}
\end{figure}

\begin{figure}[hbtp]
\centerline{
\epsfxsize=90.mm
\epsffile{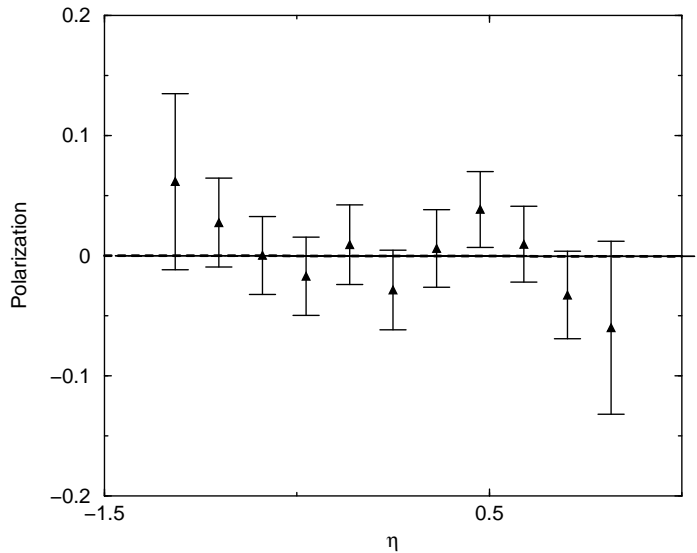}}
\caption{ Calculated polarization (dashed line)
as function of $\eta$ compared 
with the experimental data from \cite{RH1}.}
\end{figure}

It is known that in high energy $p-A$ collisions \cite{bu}, the
$p\rightarrow \Lambda$ process, produces polarized $\Lambda$ hyperons.
This result may be explained in terms of a quark exchange, of the type
$u\rightarrow s$, where an $u$ quark of the incoming proton is exchanged by
a $s$ quark, and this reaction leads to significant polarization, transversal
to the reaction plane.
In A-A colissions this effect is not expected to occur. As pointed by  Panagiotou in 1986 
\cite{pan}, a vanishing polarization should be considered as a sign of quark-gluon plasma 
formation. In \cite{pj}-\cite{lw2} the polarization of such reactions is well studied, and
it has been shown that a very small polarization is expected.
As it was shown in \cite{gyu}, if we consider a polarized hadron, produced in 
the interior of a quark-gluon plasma, observing that the mean free path
for these particles is considerably high (in A-A collisions), the effect of
successive rescattering is to attenuate the effect of this polarization and
even the information of the initial plane of production is lost. Interesting ideas 
about these processes may be found in \cite{bec1}, \cite{bec2}.  
So, according to the models above, in high energy collisions, the hyperons are produced unpolarized, and is expected 
that this polarization becomes smaller after the rescattering. But what should happen  
if the final-state interactions are considered?

Final-state interactions is a kind of effect that is very important in 
many systems, as for example, in the study of $CP$ violation in non-leptonic
hyperon decays, where the final amplitude is determined
 by the amplitude
resulting from the final-state strong interactions \cite{ccb}, \cite{Kam}, 
\cite{ta1}. As we have shown \cite{cy}, this effect is fundamental in the understanding of the polarization
of anti-hyperons in $pA$ collisions.
So, it is 
very reasonable to think that something similar might occur in heavy-ion
collisions, where the systems are very large, and the energy, very high.
In a large system, such as a RHIC collision,the probability of final interactions increase, and
this effect becomes more important. The question is if a unpolarized produced $\Lambda$, may become polarized, 
after the final interactions.

 As we verified, in p-A collisions \cite{cy},
significant polarization may be obtained when unpolarized particles
interact near the surface,
 as for example
in the $\overline{\Xi}^+$ production. In this paper, performing the calculations, we have shown that the final 
polarization remains very small (almost negligible), and this fact is due to  two reasons. The first one, is that 
the asymmetry in the polarization occurs due to the asymmetry of the
system, what is determined by the parameter $\beta$, that shows the
shape of the rapidity distribution. For large values of $\beta$ 
($\sim$2-3, that appears in p-A collisions), in the forward direction this
distribution is sharp, and polarization occurs. In A-A collisions, 
as we have shown, $\beta$ is
very small ($\beta$=0.14 for the data studied in this paper), so the
distribution is smooth, what determines cancelation of the final polarization. 
The second reason, is that the $\Lambda$ polarization in the $\pi\Lambda$ interaction
is not large (see Figure 4), and when the average is calculated, it is  almost totally 
washed out. 
So, the mechanism that is responsable for the polarization in p-A collisions,
in high energy A-A collisions has exactly the opposite effect, and destroys
most of the signs of polarization. We must remark 
the consistency of the hydrodynamical approach for
these collisions that works for p-A and for A-A collisions.

A final question is if a particle that obtains a large polarization in the final interactions 
(as for example $\Xi$ or $\Sigma$, see \cite{cy}) may be observed with some polarization in
the RHIC systems. This question shall be discussed in a next work.




\end{document}